# Evaluating sugarcane bagasse-based biochar as an economically viable catalyst for agricultural and environmental advancement in Brazil through scenario-based economic modeling

Sebastian G. Nosenzo

*PSR Energy Consulting and Analytics, Praia de Botafogo, 370, Botafogo, Rio de Janeiro, RJ, 22250-040, Brazil*

ABSTRACT

The increasing global demand for sustainable agricultural practices and effective waste management has highlighted the potential of biochar as a multifaceted solution. This study evaluates the economic viability of sugarcane bagasse-based biochar in Brazil, focusing on its potential to enhance agricultural productivity and contribute to environmental sustainability. While existing literature predominantly explores the production, crop yield benefits, and carbon sequestration capabilities of biochar, there is a notable gap in comprehensive economic modeling and viability analysis for the region. This paper aims to fill this gap by employing a scenario-based economic modeling approach, incorporating relevant economic models. Findings include that biochar implementation can be economically viable for medium and large sugarcane farms (20000-50000 hectares) given the availability of funding, breaking even in about 7.5 years with an internal rate of return of 18% on average. For small farms, biochar can only be viable when applying biochar to the soil, which in all scenarios is found to be the more profitable practice by a large margin. Sensitivity analyses found that generally, biochar becomes economically feasible at biochar carbon credit prices above $120 USD/tCO$_2$e, and at sugarcane bagasse availability percentages above 60%. While the economic models are well-grounded in existing literature, the production of biochar at the studied scales is not yet widespread, especially in Brazil and uncertainties can result. Reviewing the results, the land application scenario was found to be the most viable, and large farms saw the best results, highlighting the importance of scale to biochar operation. Small and medium farms with no land application were concluded to have no and questionable viability, respectively. Overall, sugarcane bagasse-based biochar can be economically viable, under the right circumstances, for agricultural and environmental advancement in Brazil.

## 1. Introduction

Among the wide array of increasingly relevant solutions to waste management and climate change, biochar emerges as one that can tackle multiple significant problems at once, and it comes with a host of co-benefits. Dubbed by the World Economic Forum as "carbon removal's 'jack of all trades'" (World Economic Forum, 2023), recent literature on the material suggests that biochar holds the key to the most important climate and agricultural solutions of today.

In Brazil, biochar's potential is immense, due to the country's large agricultural sector. While there are sizable bodies of research on biochar's soil amendment and carbon sequestration capabilities, as well as the environmental effect of the material, studies on the economic feasibility of biochar production are fewer. There exist a number of economic analyses of biochar in locations around the world such as Canada, Australia, China, and the USA, but such research for Brazil does not exist in detail and depth. With the goal of evaluating sugarcane bagasse-based biochar as an economically viable catalyst for agricultural and environmental advancement in Brazil, this paper presents an economic analysis through scenario-based modeling.

### 1.1 Background

Biochar is one the three products of the thermochemical process pyrolysis, in which biomass materials are converted at high temperatures, and in the absence of an oxidizing agent, such as O2, into liquids (bio-oil), non-condensable gasses (biogas), and solids (biochar) (Almeida et al., 2022). The process works by breaking down the large structural biomass molecules of the feedstock (Ramirez et al., 2019), and is able to do so utilizing practically all biomass material, unprocessed or processed (Parmar et al., 2014) (Patel et al., 2017).

Of particular interest, however, is the use of agricultural crop residues as the feedstock for pyrolysis. This is due to the fact that doing so would address the buildup of the crop residues left on a farm after harvest, reducing or eliminating residue disposal costs (Miranda et al., 2021), or otherwise finding utility for the residues that would release 70% of its carbon content into the atmosphere within one year (Souza et al., 2017) (Ahmed et al., 2016) (Cardoso et al., 2013).

The search in recent times for alternatives to fossil fuels has resulted in a wide range of prospective technologies, and crop residues have been implemented in various biomass energy structures. However, pyrolysis stands out from these alternative biomass conversion processes in that on top of valorizing waste streams (López, 2024), it brings many unique advantages such as different biomass sources application, reducing the competition between fuel and food by using unwanted material, and significant energetic optimization efficiencies (Miranda et al., 2021) (Gonçalves et al., 2017).

*1.2 Biochar's Properties*

The properties of biochar can be categorized into three types: soil enhancement, crop yield increase, and carbon sequestration. While soil enhancement and crop yield increase are closely correlated, the two factors present distinct benefits.

Biochar's properties as a soil amendment are in part what makes it such a highly appealing material to convert biomass into. The term "biochar" was developed to refer to the substance previously called "Terra Preta de Índio", a particularly fertile soil discovered near the ruins of a pre-Columbian civilization located in the Amazon basin, contrasting with the typically nutrient deficient soils of the Amazon Rainforest (Ahmed et al., 2016) (Tenenbaum, 2009). This "Terra Preta" was used by the region's inhabitants to improve the local soil for agricultural purposes (Lehmann et al., 2003). Research in the centuries following its discovery has unearthed an array of benefits that biochar poses when applied to soil, especially soils with existing deficiencies.

Biochar application has been found to increase microbial activity in the soil, as well as accelerate the mineralization of organic matter, nutrient cycling, and organic matter content, resulting in higher levels of nutrient uptake by a plant grown in that soil (Joseph et al., 2021) (Li et al., 2024) (Sohi, 2012) (Zafeer et al., 2024) (Hermans, 2013) (Yu et al., 2018) (Voorde, 2014). Furthermore, biochar can affect crop plant root morphology by promoting fine root proliferation, increasing specific root length, and decreasing both root diameter and root tissue density (Purakayastha et al., 2019). In addition to these properties, biochar improves soil pH and soil structure in terms of porosity (50% increase) and bulk density (40% decrease) depending on soil type (Galinatio et al., 2011) (Chang et al., 2021), improving aeration and water content up to over 100%, an aspect especially beneficial in arid regions (Blanco-Canqui, 2017). For agricultural purposes, this results in less water use and reduced frequency of irrigation, tillage, and other crop management practices.

As a result and in addition to these soil amendment properties of biochar, its implementation in agriculture results in relevantly increased crop yields (Schmidt et al., 2021). Depending on soil and crop, increases with biochar implementation resulted in 10% to 30% crop yield, with 15% on average (Joseph et al., 2021). This metric describing a 15% increase in crop yield over traditional fertilizers is commonly found with biochar application with various crops such as corn, wheat, rice, and soybean (Borges et al., 2020) (Campion et al., 2023) (Lima and White, 2017) (Ahmed et al., 2016), with the combined use of biochar and fertilizer presenting even higher increases in some studies (Brown et al.).

As evidenced in a study by Borges et al. conducted in Brazil, biochar implementation resulted in a 15% higher crop yield compared to traditional triple superphosphate fertilizers. Additionally, the study noted that biochar maintained a soil pH between 7 and 8, which enhanced phosphorus availability, further benefiting crop growth. This highlights biochar's ability not only to replace but, in some cases, surpass the effects of conventional fertilizers, particularly in tropical and subtropical regions.

Importantly, studies found that biochar land application has a much greater crop yield increase effect in tropical climates over temperate ones, with averages up to 25% increases in crop yield in tropical agriculture (Jeffery et al., 2017). Brazil, being a country that experiences tropical and subtropical climates (World Bank Climate Change Knowledge Portal, 2021) fits this criteria and holds a further potential in its agriculture for biochar implementation.

The most relevant aspect of biochar is its carbon sequestration property. Through the process of pyrolysis, biochar retains the majority (60-80%) of the carbon content of the biomass feedstock (Martins et al., 2021), resulting in a stable solid, rich in carbon and capable of resisting chemical and microbial breakdown (USDA) (Brown et al.). When applied to the soil, which, as mentioned, comes with a multitude of soil enhancement and crop yield increase benefits, the material effectively sequesters the carbon, maintaining the carbon content in its stable form for hundreds to thousands of years (FAO). This, when done according to the approved biochar carbon credit methodologies and verified by certification agencies, generates carbon credits for the removal of carbon, which can be a significant stream of revenue for the entity producing and applying biochar. The removal credits generated via biochar land application are frequently regarded as superior to alternative reduction credits in the global carbon market, commanding much higher prices. This is because removal efforts are often easier to justify and measure than reduction projects. According to recent forecasts, demand for the project type may expand 20 times over the next decade, and evidence suggests that the relevant market size is increasing steadily (MSCI, 2024) (Precedence Research, 2023). In fact, recent purchases of biochar credits by corporate giants such as Microsoft and JP Morgan Chase points to the growing significance and confidence in the biochar credit market (MSCI, 2024). Currently, biochar carbon credits sales have prices that range between 100 and 200 $ USD per metric ton of $CO_2$ removed, with the average price being $179/tCO$_2$e (World Economic Forum, 2023) (López, 2024).

It is important to consider, however, that biochar is a relatively new player on the global scene and these benefits might not see their full potential for years. For example, the worldwide biochar demand is still almost 500 times smaller than the fertilizer demand, even though biochar can produce better results (Campion et al., 2023). On top of this, a combination of a lack of standardization and a lack of scale in the industry may be holding biochar back in terms of widespread use (Meadows, 2015) (López, 2024) (Campion et al., 2023). Overall and in time, though, with the global crop residue production increasing 33% in ten years, crop residue-based biochar production could be an effective tool to combat greenhouse gas emissions and invigorate agricultural industries simultaneously through the use of biochar. (Brown et al.) (Borges et al., 2020) (Cherubin et al., 2018).

*1.3 Biochar in Sustainable Agriculture*

The global shift toward sustainable agriculture is driven by the need to reconcile growing food demands with the imperative to minimize environmental degradation. Agriculture is a major contributor to greenhouse gas emissions, soil depletion, and biodiversity loss, making it essential to adopt practices that not only enhance productivity but also restore ecological balance. Biochar, with its unique combination of benefits, is considered a highly promising solution. By converting agricultural waste into biochar, the dual objectives of reducing waste and enhancing the resilience of farming systems can be achieved, positioning biochar as a key player in sustainable agriculture.

In Brazil, where agriculture forms a substantial part of the economy, the potential for biochar application is significant. As one of the largest agricultural producers, the country faces unique challenges in balancing its agricultural expansion with environmental conservation. Monoculture practices, deforestation, and soil degradation are significant concerns, particularly in the production of crops like sugarcane. Biochar's ability to repurpose agricultural residues such as sugarcane bagasse offers a practical and scalable solution.

The integration of biochar into Brazil's agricultural systems could provide multiple co-benefits: reducing waste, improving soil health, and generating additional revenue through carbon credits.

*1.4 Brazilian Sugarcane Production*

While biochar production can be executed using any biomass or crop residue, sugarcane bagasse is an especially appealing choice for a feedstock due to its high content of organic compounds cellulose and hemicellulose vital to the pyrolysis process, its positive influences of the final biochar's soil amending properties, and, in countries where it is a main crop, its abundance.

With sugarcane as one of its main commodities, Brazil is the world's largest producer of the crop, producing over 700 million tons and 25% of the world production (Miranda et al., 2021) (OECD-FAO, 2022). In Brazil and in the main developing countries, sugarcane plays an important role in the energy and economic systems, producing the large-scale products sugar and ethanol (Clauser et al., 2018).

In the context of biochar, the main part of sugarcane that is used for pyrolysis is the bagasse that remains after the sugarcane's use for sugar and ethanol production. On average 1 ton of sugarcane produces 280 kg of bagasse (Almeida et al., 2022) (Zafeer et al., 2024) (Cardona et al., 2010).

This sugarcane residue can be burned in boilers for power generation, but this is still underused by the vast amount available and much of the bagasse's potential remains largely untapped (Miranda et al., 2021) (Cherubin et al., 2018) (Lima and White, 2017). Moreover, it is established that burning crop residues greatly harms human health and the environment by releasing greenhouse gasses into the atmosphere (Nematian et al., 2021). Thus, the alternative of using the bagasse for biochar production is even more appealing for Brazil.

*1.5 Sugarcane Bagasse-Based Biochar*

Such use of sugarcane bagasse to make sugarcane bagasse-based biochar would involve the aforementioned process of pyrolysis. The process can yield differing results and products (liquids, gasses, and solids in varying amounts (Brownsort, 2009)) depending on a multitude of factors, of which the type of pyrolysis is a main aspect. Multiple studies have found that the so-called "slow pyrolysis" at 300-500 ºC not only results in the highest amount of biochar (50.3%), but is also more cost effective (Almeida et al., 2022) (Homagain et al., 2015) (Kung et al., 2013). Overall, comparisons of the pyrolysis biochar system with other bioenergy production systems for carbon abatement found that the pyrolysis biochar system is 33 % more efficient than direct combustion, even if the soil amendment benefits of biochar are ignored (Hammond et al. 2011).

In the context of this study, which considers biochar in Brazilian sugarcane production, the liquid and gas products of pyrolysis, which would have to undergo refining, treatment, and transportation (Almeida et al., 2022), incurring costs, are disregarded and assumed to be utilized by the producer as fuel to start the pyrolysis or for heating. Finally, questions remain as to biochar's economic sustainability. This study aims to provide a comprehensive evaluation of sugarcane bagasse-based biochar's economic viability in Brazil as a protagonist in the country's agricultural and environmental initiatives.

**2. Methodology**

To assess the economic viability of sugarcane bagasse-based biochar as a catalyst for agricultural and environmental advancement in Brazil, various economic models were implemented. These include a life cycle cost analysis, cost-revenue-based break even analysis, return on investment, cost-benefit analysis, net present value and internal rate of return, and sensitivity analyses, all forming a comprehensive Biochar Economic Viability Model.

*2.1 Study Area, Scenarios, and Values*

In order to evaluate sugarcane bagasse-based biochar as an economically viable catalyst for agricultural and environmental advancement in Brazil, and to analyze biochar economically within an area where its aforementioned significant potential exists, scenarios within the study area of Brazil were considered. Scenarios A–the direct sale of biochar (carbon credits) following production–and B–the land application of produced biochar followed by the sale of surplus biochar (carbon credits–were chosen to reflect possible business structures adopted by a farm producing biochar. Each scenario (A, B) was considered for farms of 10000 ha, 20000 ha, and 50000 ha in size, for a total of 6 distinct scenarios.

Table 1 and Table 2 detail relevant values determined through a review of existing biochar-related literature and relevant values determined for this study, respectively. All chosen values or assumptions for this study were developed with the cooperation and insights from PSR Energy Consulting and Analytics in Rio de Janeiro, Brazil.

**Table 1.** Values determined through a review of existing biochar-related literature

| Description | Val. | Unit | Reference |
|---|---|---|---|
| Sugarcane yield | 73.7 | t ha$^{-1}$ | (OECD-FAO, 2022) |
| Bagasse:sugarcane | 28 | % | (Cardona et al., 2010) |
| Biochar:bagasse | 50.3 | % | (Almeida et al., 2022) |
| Dry:bagasse[1] | 60 | % | (Almeida et al., 2022) |
| Land app. rate | 4.2 | t/ha | (Lefebvre et al., 2020) |
| Biochar yield inc. | 15 | % | (Borges et al., 2020) |
| Interest rate | 9.00 | % | (Banco do Brasil, 2024) |
| Inflation rate | 3.00 | % | (Statista, 2024) |
| BRL to USD | 0.18 | $/R$ | (July 2024) |
| Sugarcane m. p. | 30.6 | $ t$^{-1}$ | (USDA, 2024) |
| Carbon credit m. p. | 179 | $[2] | (WEF, 2023) |
| C.C.[3] issuance fee | 0.30 | $[4] | (Gold Standard, 2023) |
| C.C. review fee | 1.9 | k$ | (Gold Standard, 2023) |
| Fertilizer cost | 34 | $ ha$^{-1}$ | (Amorim et. al, 2022) |
| Crop mgmt. cost | 437 | $ ha$^{-1}$ | (Amorim et. al, 2022) |
| Land app. cost[5] | 320 | $ ha$^{-1}$ | (Amorim et. al, 2022) |

**Table 2.** Values chosen for this study's economic modeling

| Description | Value | Unit |
|---|---|---|
| Analysis time frame | 20 | y |
| Farm size (small) | 10000 | ha |
| Farm size (medium) | 20000 | ha |
| Farm size (large) | 50000 | ha |
| Planning/design cost % of init. investment | 5 | % |
| Permit cost % of init. investment | 2 | % |
| Maintenance cost % of FCI[6] | 2 | % |
| Industrial scale factor | 0.7 | x |

[1]Bagasse needs to be dried from 50% to 10%wt moisture for pyrolysis [2]$ tCO$_2$e$^{-1}$ [3]Carbon Credit [4]$ credit$^{-1}$ [5]Crop management cost, including tillage, irrigation, and other soil preparation [6]Based on (Ramirez et al., 2019)

*2.2 Life Cycle Cost Analysis*

To calculate the life cycle costs for each studied scenario, the following model was used (Eq. 1-3)

$$C_{total} = C_i + \sum_{y=0}^{n} C_y \tag{1}$$

$$C_i = C_{pd} + C_{iw} + C_{es} + C_{pv} \tag{2}$$

$$C_y = C_m + C_o + C_l + C_{cc} + C_{la} \tag{3}$$

where $C_{total}$ is total life cycle costs, $C_i$ is initial investment costs, $C_y$ is yearly costs, $n$ is the total number of years $C_{pd}$ is planning and design costs, $C_{iw}$ is indirect and working capital costs, $C_{es}$ is equipment and setup costs, $C_{pv}$ is permit and carbon credit project validation costs, $C_m$ is maintenance costs, $C_o$ is operation costs, $C_l$ is labor costs, $C_{cc}$ is carbon credit certification costs, and $C_{la}$ is land application costs, for which Scenario A has none.

To calculate the value of the variables, a reference study of three biofuel plants simulating an 84,000 t/y use of sugarcane bagasse in Queensland, Australia was used (Ramirez et al., 2019). Values were then location-adjusted to be relevant for the study area of Brazil, using a ratio of the Price Level Index (PLI) of both locations for the equipment, setup, and operation costs (Statista, 2022), and a ratio of the minimum wage (2024) of both locations for the labor costs (Câmara dos Deputados, 2023) (Australian Government, 2022). Reference values from the (Ramirez et al., 2019) study are outlined in Table 3.

**Table 3.** Reference cost values in $ USD millions from (Ramirez et al., 2019) for life cycle cost analysis

| Description | Value | Unit |
|---|---|---|
| Total installed costs | 38.42 | $M |
| Total indirect + working capital costs | 13.63 | $M |
| Labor cost | 1.17 | $M/y |
| Operation cost excl. feedstock | 2.08 | $M/y |

*2.3 Cost-Revenue-Based Break Even Analysis*

To conduct a cost-revenue-based break even analysis for each studied scenario, cumulative life cycle costs from the life cycle cost analysis were used, as well as life cycle revenues calculated using the following models (Eq. 4) and (Eq. 5) for Scenario A and Scenario B, respectively

$$R_{total} = R_b + R_{cc} \tag{4}$$

$$R_{total} = R_b + R_{cc} + R_s \quad (5)$$

where $R_{total}$ is total life cycle revenues, $R_b$ is revenues from biochar sales, $R_{cc}$ is revenues from carbon credit sales, and $R_s$ is revenues from increased sugarcane sales, for which Scenario A has none.

*2.4 Return on Investment*

To calculate the return on investment percentage per year for each studied scenario, the following model was used (Eq. 6)

$$ROI = \frac{R_t - C_t}{C_i} \times 100 \quad (6)$$

where $R_t$ is total revenues for year $t$, $C_t$ is the total costs for year $t$ (where $C_t$ for the first year does not include the initial investment cost), and $C_i$ is the cost of the initial investment.

*2.5 Cost-Benefit Analysis*

To compare the costs and benefits for each studied scenario, total life cycle costs from the life cycle cost analysis were used. In addition, to calculate the benefits, the following model was used (Eq. 7-8)

$$B_{total} = R_{total} + S_{total} \quad (7)$$

$$S_{total} = S_f + S_o \quad (8)$$

where $B_{total}$ is total benefits, $R_{total}$ is total life cycle revenues from the cost-revenue-based break even analysis, $S_{total}$ is total savings, applicable to Scenario B only, as these savings are results of biochar land application, $S_f$ is fertilizer savings, and $S_o$ is operational savings.

To then conduct the cost-benefit analysis, the following model was used (Eq. 9)

$$BCD = B_{total} - C_{total} \quad (9)$$

where $BCD$ is benefit-cost difference, $B_{total}$ is total benefits, and $C_{total}$ is total costs.

*2.6 Net Present Value and Internal Rate of Return*

To calculate the net present value, or the present value of cash flows over the analysis time frame, the following model was used (Eq. 10)

$$NPV = \sum_{t=0}^{n} \frac{CF_t}{(1+r)^t} \quad (10)$$

where $NPV$ is net present value, $n$ is the total number of years, $CF_t$ is total cash flow for year $t$, $t$ is the year, and $r$ is the discount rate, which is used to translate future cash flows into present value by considering factors such as expected inflation and the ability to earn interest (EPA, 2021). It is equal to the Brazilian nominal interest rate which accounts for both interest and inflation, and was determined using current forecasts for the years following 2024 (Banco do Brasil, 2024) (Bloomberg, 2024) (Brazilian Report, 2023).

To calculate the internal rate of return, or the annual rate of growth that an investment is expected to generate, the following model was used (Eq. 11)

$$0 = \sum_{t=0}^{n} \frac{CF_t}{(1+IRR)^t} \quad (11)$$

where $IRR$ is internal rate of return, $n$ is the total number of years, $CF_t$ is total cash flow for year $t$, and $t$ is the year.

*2.7 Sensitivity Analysis*

The variables of (1) carbon credit market price and (2) available bagasse were identified as relevant variables to the results of the economic viability analysis of sugarcane bagasse-based biochar.

For the variable of carbon credit market price, current metrics suggest that biochar carbon credits are worth between $100 and $200 per ton of $CO_2$ emissions reduced/sequestered, with an average market price of $179/t$CO_2$e (López, 2024) (World Economic Forum, 2023). However, these prices are a reflection of a relatively new market (Carbon Credits, 2024), and at the scale of carbon credits in this study, this price may be expected to not hold. Therefore, a sensitivity

analysis with their market price ranging from $50 to $200 per ton of $CO_2$e was chosen.

For the variable of bagasse availability, current sugarcane farms in Brazil are commonly using a percentage of their produced sugarcane bagasse for bioelectricity or heat in sugar and ethanol production processes (Almeida et al., 2022). The leftover bagasse (ex. 70% if the farm utilizes 30% for bioelectricity, a common practice), is what is available to be converted into biochar through pyrolysis without affecting current costs. Since the use of bagasse for bioelectricity varies from farm to farm depending on production, equipment, and bioelectricity usage, the percentage of bagasse available can also vary. Therefore, a sensitivity analysis with the percentage of available bagasse ranging from 50% to 90% was chosen.

As such, sensitivity analyses were conducted for both of these variables, analyzing the net present value of all studied scenarios with the variables ranging from 50 to 200 ($/t$CO_2$e) for carbon credit market price and 50 to 90 (%) for bagasse availability.

*2.8 Selection and Implementation of Models*

The economic models employed in this study were selected to provide a comprehensive assessment of biochar's financial viability in Brazil, where both agricultural and carbon markets present unique challenges and opportunities. A life cycle cost analysis (LCCA) and net present value (NPV) were chosen to capture the full scope of investment and operational costs over the long term, particularly suited for high upfront capital projects like biochar production. To complement these, the internal rate of return (IRR) and cost-benefit analysis (CBA) models were applied to provide a clearer picture of profitability under different farm sizes and operational strategies. These models are well-suited for projects with high initial capital and long-term revenue streams, as they capture the full scope of investment, operating costs, and potential returns over time.

Given the nascent nature of the biochar and carbon credit markets, scenario-based modeling with these tools were necessary for a detailed examination of different operational scales and conditions.

To ensure scientific rigor and reproducibility, this study implemented these models using inputs derived from peer-reviewed sources, industry consultations, and localized agricultural data. Key parameters—such as bagasse conversion rates, carbon credit pricing, and operational costs—were adapted to reflect the specific conditions of the Brazilian sugarcane industry.

A series of sensitivity analyses on the most uncertain factors like carbon credit market prices and bagasse availability allowed for an exploration of the effects of market fluctuations on profitability and ensured that the results were adaptable to future shifts in both agricultural and environmental policy.

By integrating these established models and adapting them to Brazil's specific agricultural context, the study ensures a rigorous, replicable approach, making this analysis both scientifically grounded and highly relevant for evaluating biochar's feasibility in various agricultural settings.

**3. Results and Discussion**

*3.1 Life Cycle Cost Analysis*

The life cycle costs of all six studied scenarios are detailed in Fig. 1.

The initial investment for both scenarios of the small farm amounted to $57.5 million USD, with yearly costs over 20 years around $91 million USD for Scenario A and $148 million USD for Scenario B. For the initial investment, the highest amount lies in the equipment and setup of the pyrolysis plant, totaling $39.5 million USD. Out of the yearly costs, the highest costs are operation, followed by maintenance and labor for Scenario A, at $61 million USD, $23 million USD, and $7 million USD, respectively over 20 years accounting for inflation, and operation followed by land application and maintenance for Scenario B, at $61 million USD, $57 million USD, and $23 million USD, respectively over 20 years accounting for inflation.

For both scenarios of the medium farm, the initial investment amounted to $93 million USD, with yearly costs over 20 years around $148 million USD for Scenario A and $263 million USD for Scenario B. The equipment and setup of the pyrolysis plant

account for the largest portion of the initial investment, coming in at $64 million. Once again, out of the yearly costs, the highest costs are operation, followed by maintenance and labor for Scenario A, at $99 million USD, $37 million USD, and $11 million USD, respectively over 20 years accounting for inflation, but for Scenario B, the highest cost was land application, followed by operation and maintenance, at $115 million USD, $99 million USD, and $37 million USD, respectively over 20 years accounting for inflation.

For the large farm, the initial investment in both scenarios was $177 million USD. Yearly costs over the 20 years were $281 million USD for Scenario A and $568 million USD for Scenario B. For the initial investment, the highest amount was the equipment and setup of the pyrolysis plant, totaling $121 million USD. Over 20 years and accounting for inflation, total yearly costs are most expensive for operation, followed by maintenance and labor in Scenario A, coming in at at $189 million USD, $70 million USD, and $21 million USD, respectively, and land application followed by operation and maintenance for Scenario B, coming in at $287 million USD, $189 million USD, and $70 million USD, respectively.

Overall, operation and equipment pose the largest costs, with land application costs being among the most expensive for all farms in Scenario B. Planning & design, permit, and carbon credit certification costs are negligible in all scenarios, accounting for less than 10% of total costs combined. Comparing scenarios directly, scenarios for the medium farm cost about 2 times the corresponding small farm scenarios, and large farm scenarios cost twice their corresponding medium farm scenarios, and four times their corresponding small farm scenarios. In general, Scenario B costs around one and a half times Scenario A. Finally, all scenarios would require a significant initial investment as well as significant yearly costs. The economic feasibility of such scenarios depends on available investment capital as well as the resulting revenue and savings.

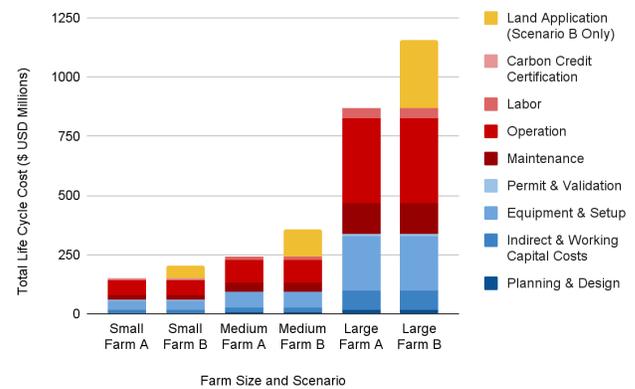

**Fig. 1.** Life cycle cost components in $ USD millions over 20 years for studied scenarios

*3.2 Cost-Revenue-Based Break Even Analysis*

Fig. 2 details the break even for the studied scenarios based on cumulative costs and revenues, while Table 4 details the respective break even values in years.

The revenue to cost ratio over 20 for each scenario (increasing farm size, Scenario A then B), the revenue to cost ratio was 1.7, 2.3, 2.1, 2.7, 2.7, and 3.2, respectively, with large farm scenarios being the most profitable overall. All investments took less than 13 years to break even, and the large farm in Scenario B broke even in 5.15 years.

*3.3 Return on Investment*

Fig. 3 details the return on investment percentages for the studied scenarios, and Table 5 details the respective total return on investment percentages.

For each scenario (increasing farm size, Scenario A then B), the return on investment percentages ranged from 2 to 33%, -1 to 66%, 4 to 43%, 0 to 83%, 7 to 59%, and 1 to 113% over the 20 years, respectively

The average return on investment per year for each scenario or each scenario (increasing farm size, Scenario A then B) was 13%, 27%, 18%, 35%, 26%, and 48%, respectively. Scenario B farms did significantly better than their Scenario A counterparts in terms of return on investment for all farm sizes

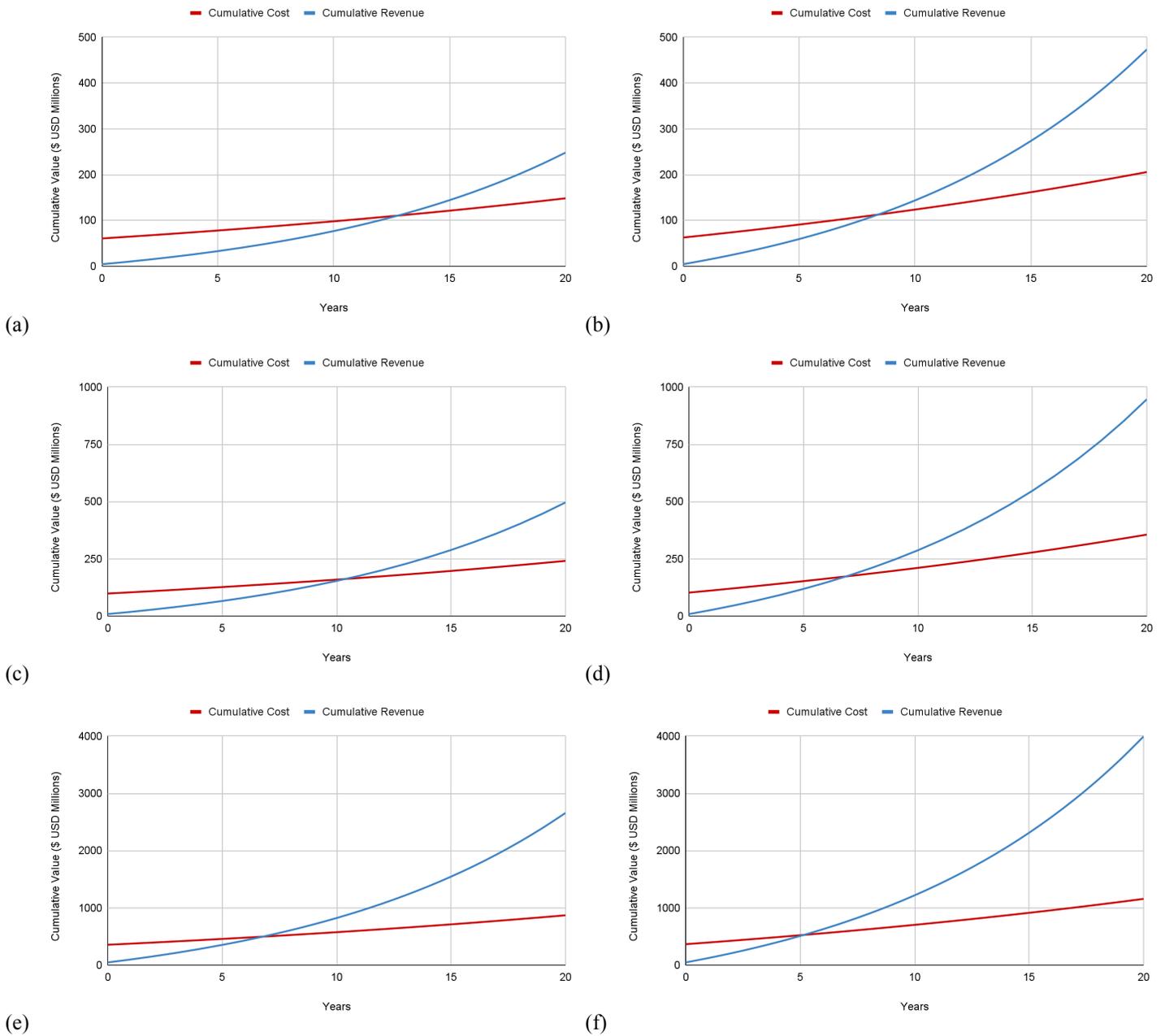

**Fig. 2.** Life cycle break-even analysis based on cumulative costs and revenues in $ USD millions over 20 years for a 10000 ha farm in Scenario A (a), 10000 ha farm in Scenario B (b), 20000 ha farm in Scenario A (c), 20000 ha farm in Scenario B (d), 50000 ha farm in Scenario A (e), and 50000 ha farm in Scenario B (f)

**Table 4.** Break-even year values for studied scenarios

| Farm Size | Scenario | Years |
| --- | --- | --- |
| Small (10000 ha) | A (Biochar Sale) | 12.77 |
| Small (10000 ha) | B (With Land Application) | 8.44 |
| Medium (20000 ha) | A (Biochar Sale) | 10.46 |
| Medium (20000 ha) | B (With Land Application) | 6.85 |
| Large (50000 ha) | A (Biochar Sale) | 7.78 |
| Large (50000 ha) | B (With Land Application) | 5.15 |

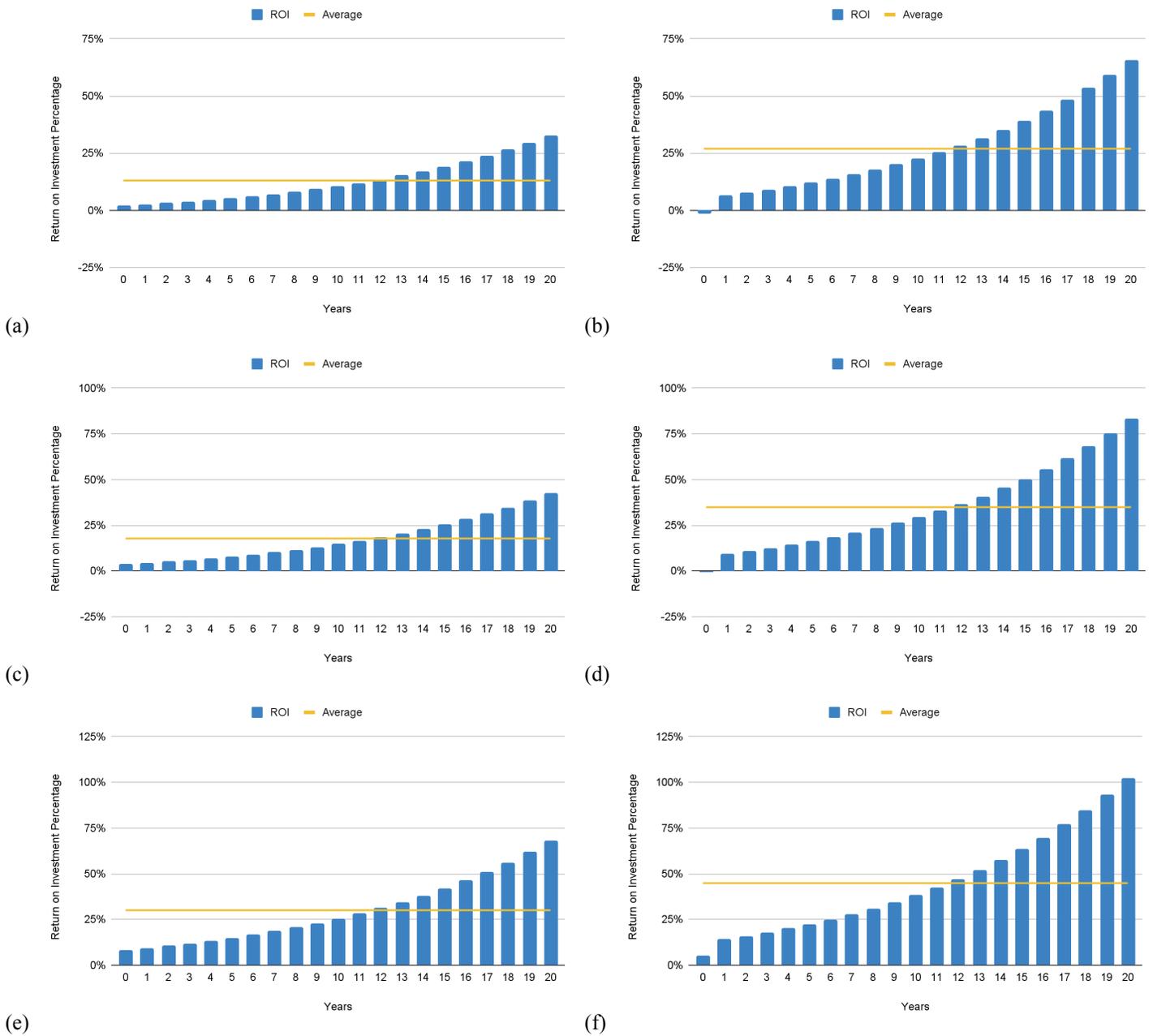

**Fig. 3.** Return on investment percentages per year over 20 years for a 10000 ha farm in Scenario A (a), 10000 ha farm in Scenario B (b), 20000 ha farm in Scenario A (c), 20000 ha farm in Scenario B (d), 50000 ha farm in Scenario A (e), and 50000 ha farm in Scenario B (f)

**Table 5.** Total return on investment percentage over 20 years for studied scenarios

| Farm Size | Scenario | ROI Percentage |
| --- | --- | --- |
| Small (10000 ha) | A (Biochar Sale) | 274 |
| Small (10000 ha) | B (With Land Application) | 566 |
| Medium (20000 ha) | A (Biochar Sale) | 373 |
| Medium (20000 ha) | B (With Land Application) | 733 |
| Large (50000 ha) | A (Biochar Sale) | 542 |
| Large (50000 ha) | B (With Land Application) | 1015 |

## 3.4 Cost-Benefit Analysis

The cost benefit comparison for the studied scenarios is detailed in Fig. 4.

The difference between the total revenues and total cost for all scenarios was positive, and for each one (increasing farm size, Scenario A then B), came out to 100, 408, 255, 873, 782, and 2326 million $ USD, respectively. Again, Scenario B farms outperform their Scenario A counterparts for all farm sizes, and large farm scenarios are the highest performing.

It can be noted that the fertilizer and crop management savings that result from biochar land application for Scenario B farms considered as a benefit in the analysis greatly exceed the land application cost that Scenario B farms spend.

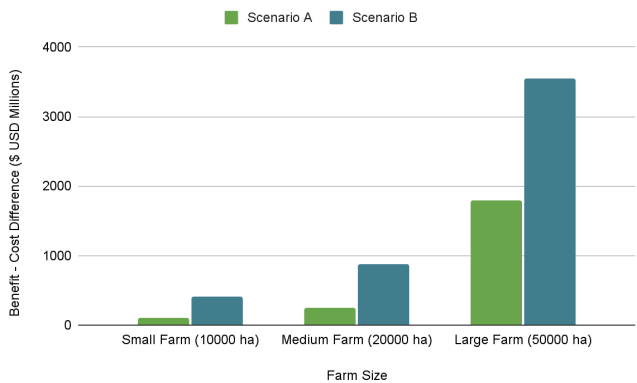

**Fig. 4.** Cost-benefit comparison in $ USD millions over 20 years for studied scenarios

## 3.5 Net Present Value and Internal Rate of Return

Fig. 5 details the net present value of 20 years for each of the studied scenarios.

The net present value for all scenarios except for small farm Scenario A is positive, indicating a rate of return above the discount rate, and, in general, a positive investment. For small farm Scenario A, the net present value is negative, indicating a negative investment and rendering the biochar implementation in the scenario not economically viable for the purposes of this study's evaluation. For each scenario, (increasing farm size, Scenario A then B), the net present value is -6, 50, 25, 136, 158, and 436 million $ USD, respectively. Scenario B farms, due to their increased revenue streams, had a higher net present value than their Scenario A counterparts, and large farm scenarios had net present values about 5 times greater than the medium farms, and over 8 times greater than the small farms depending on the scenario.

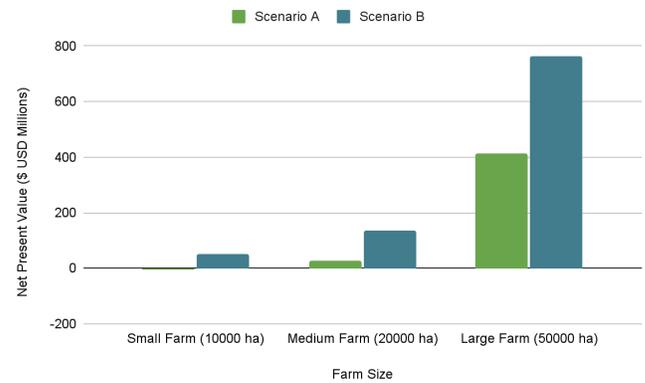

**Fig. 5.** Net present value in $ USD millions over 20 years for studied scenarios

The internal rate of return over 20 years for all studied scenarios are detailed in Fig. 6.

The internal rate of return, indicating the annual rate of growth that the investment is expected to generate, of 8%, 16%, 11%, 19%, 17%, and 25% for each scenario (increasing farm size, Scenario A then B) reveal that while all scenarios show positive growth, the larger farm scenarios demonstrate significantly higher returns. While all scenarios have a positive internal rate of return, the large farm scenarios have values around 1.6 times those of the small farm and 1.3 times those of the medium farm. Additionally, the internal rates of return for small farm Scenario A and medium farm Scenario A, being under 15%, indicate an unfavorable investment given the high investment costs found in the life cycle cost analysis.

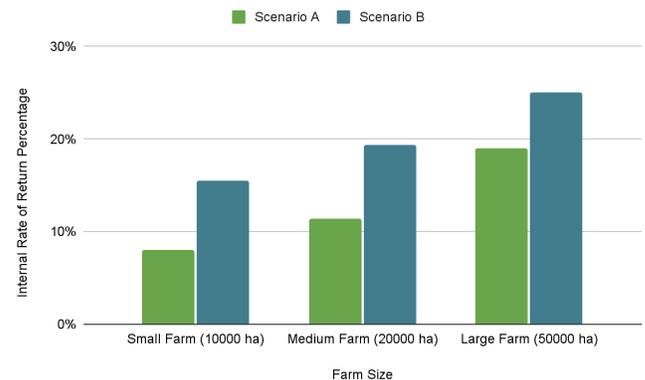

**Fig. 6.** Internal rate of return over 20 years for studied scenarios

## 3.6 Sensitivity Analysis

Fig. 7 details the sensitivity of the total net result, in present value over 20 years based on a carbon credit market price ranging from $50 to $200/tCO$_2$e.

For each scenario (increasing farm size, Scenario A then B), the carbon credit market price at which the net result was positive was $200, $100, $160, $70, $120, and $50/tCO$_2$e, respectively. In general, Scenario A for both small and medium farms produced negative results for any price below the current average market price of $179/tCO$_2$e, with small farm Scenario A not resulting in a positive result except when the price was $200/tCO$_2$e. This indicates the two scenarios may not be economically viable if the price drops. For the remaining scenarios, all produced a positive net result at about $120/tCO$_2$e, indicating an almost $60 margin for the market price to drop. This suggests that medium and large farms, particularly those in Scenario B, have a buffer to withstand fluctuations in the carbon credit market without becoming economically unfeasible. The highest performing scenario was large farm Scenario B, which had a positive net result even at $50/tCO$_2$e, a low price considering today's biochar carbon credit market. Importantly, the highest fluctuation in net result was also observed in the large farm scenarios, with results varying by almost $450 USD million depending on the price. This highlights the dependence of the economic viability of a biochar operation even in large farms on the price that the generated carbon credits can be sold for.

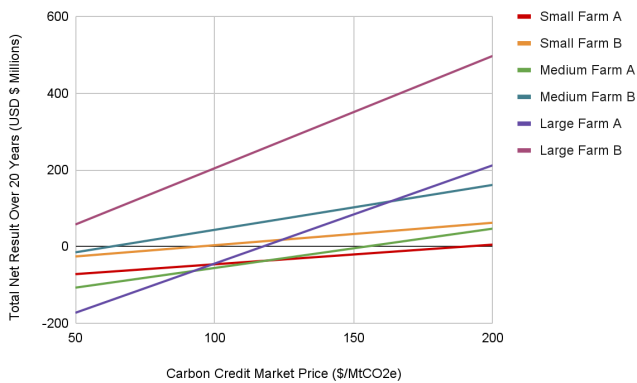

**Fig. 7.** Sensitivity of total net result over 20 years (present value) in $ USD millions based on carbon credit market prices of 50 to 200 ($/tCO$_2$e) for studied scenarios

The sensitivity of the total net result, in present value over 20 years based on bagasse availability ranging from 50% to 90% are detailed in Fig. 8.

The only scenarios to produce a negative net result at any value were small farm Scenario A and medium farm Scenario B, producing positive results at 70 % and 80% bagasse availability, respectively. For the other scenarios, the net result remains positive for bagasse availability ranging from 50% to 90%, indicating viability in most cases where bagasse availability can be affected by an individual farm's equipment and electricity demands.

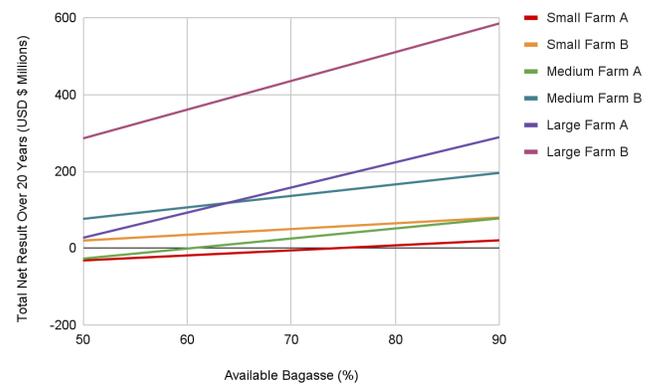

**Fig. 8.** Sensitivity of total net result over 20 years (present value) in $ USD millions based on bagasse availability of 50 to 90 (%) for studied scenarios

## 3.7 Technical and Economic Challenges

While the results of this study demonstrate the potential economic viability of biochar, particularly for large-scale operations, several technical and economic challenges remain, especially in the context of Brazil's agricultural sector. The pyrolysis process requires specialized equipment with high capital costs, making the initial investment a significant barrier, particularly for small and medium-sized farms.

Economically, the long-term feasibility of biochar production is closely tied to fluctuations in the carbon credit market and the operational costs associated with labor and energy. As demonstrated by the sensitivity analysis, carbon credit prices play a pivotal role in determining profitability. For smaller farms, biochar production becomes financially unfeasible when carbon credit prices fall. This highlights the vulnerability of biochar operations to market volatility, particularly in Brazil, where the carbon credit market is still maturing. Additionally, operational costs, particularly energy for pyrolysis and labor, remain significant, especially in regions where energy prices fluctuate or access to affordable labor is limited.

Moreover, Brazil's fragmented agricultural landscape, characterized by both large agribusinesses and smallholder farmers, makes the uniform implementation of biochar technology difficult. Smaller farms, which could benefit most from the soil-enhancing properties of biochar, are often the least equipped to handle the associated technical and economic burdens. Without targeted financial incentives or subsidies, these farms may struggle to adopt biochar on a meaningful scale.

To address these challenges, policy support is critical. Government intervention in the form of subsidies, financial incentives, or carbon credit guarantees could alleviate some of the economic pressure on smaller operations. Infrastructure development, particularly in regions where transportation costs are high, would also enhance the scalability of biochar production. Without such measures, the broad-scale implementation of biochar in Brazil remains constrained by both technical and economic limitations.

*3.8 Limitations of the Study*

This study has provided a detailed economic analysis of sugarcane bagasse-based biochar production in Brazil; however, several limitations must be acknowledged. One significant limitation is the lack of large-scale biochar operations implemented or extensively studied in practice. This introduces an inherent uncertainty regarding the real-world applicability and outcomes of the modeled scenarios. While the theoretical models and assumptions used are robust and well-grounded in existing literature, they may not fully capture the operational complexities and unforeseen challenges that might arise in actual biochar production and application.

Moreover, the values concerning the soil-amending and crop yield-increasing properties of biochar are specific to the area of the biochar operation, and significant variations in the effects of biochar application may arise depending on farm location (Li et al., 2024). Finally, the nascent nature of the biochar market, particularly in Brazil, adds another layer of uncertainty. The assumptions regarding market prices, availability of carbon credits, and economic incentives are based on current trends and projections, which can be highly variable and subject to change. While this study attempts to evaluate the impact of any uncertainties by conducting sensitivity analyses, considering these uncertainties and the evolving nature of biochar technology and market dynamics is crucial.

**4. Conclusions**

The results of this study reveal an array of conclusions on sugarcane bagasse-based biochar's economic viability. The economic models used indicate that Scenario B, which includes land application of biochar followed by the sale of surplus biochar (carbon credits), consistently outperformed Scenario A, which only considers the direct sale of biochar (carbon credits). This trend was evident across all farm sizes, with large farms demonstrating the highest economic viability. Specifically, large farm scenarios yielded the most favorable outcomes, underscoring the importance of scale to biochar's viability as a profitable catalyst for agricultural advancement in Brazil. Small farm Scenario A, however, was not economically viable, and medium farm Scenario A was highly susceptible to changes in critical values, making its economic viability questionable.

For medium and large farms, biochar production and application appeared to be economically viable under the condition that sufficient capital investment is available and that the carbon credit market prices remain stable, with some tolerance for fluctuation. The sensitivity analysis highlighted that biochar becomes feasible on average at carbon credit prices above $120 USD/tCO2e, providing a buffer against market volatility.

With increasing regulatory interest in carbon removal as a result of many nations's agreement to combat climate change in the United Nations Paris Agreement of 2015 (Xia et al., 2023), biochar's value as a capable instrument for carbon sequestration continues to increase (Nematian et al.,2021) (Galinatio et al., 2011) (Sohi, 2012). When considering the fact that the world has lost a third of its arable land due to pollution and erosion in the past 40 years (Yu et al., 2018), biochar becomes a truly interesting option for the agricultural industry. This, coupled with the scale and potential of sugarcane bagasse as a feedstock, makes sugarcane bagasse-based biochar a unique opportunity for Brazil. Regarding this possibility, key prospects for biochar to expand within Brazilian agriculture include targeted policy support, such as subsidies and carbon credit guarantees, alongside investments in agricultural infrastructure. These measures would mitigate the financial challenges smaller farms face, fostering broader adoption.

It is clear that biochar advocates will have to present a convincing argument to farmers about the benefits of biochar application in agronomy (Kulyk, 2012). Through the economic models in this study, relevant economic benefits can be observed. In addition to the revenues and savings analyzed in the cost-benefit analysis, sugarcane producers remove disposal costs and do not incur any extra transportation costs (Miranda et al., 2021), making biochar a logical addition to their business model given the funds to invest in an operation. The final verdict, therefore, lies with the specific details of an individual farm's resources, structure, and goals.

Building on the findings of this study, future research should focus on optimizing biochar production processes, particularly by improving pyrolysis efficiency to reduce energy consumption. Exploring alternative energy sources or refining operational parameters could make biochar production more cost-effective, especially for smaller farms. Additionally, refining the carbon credit certification process and ensuring stable pricing mechanisms will be crucial to biochar's long-term viability in fluctuating markets.

Further studies on region-specific economic impacts are essential, given the variability in Brazil's agricultural infrastructure and crop cycles. Tailoring biochar solutions to local conditions will enhance scalability and feasibility across diverse farming contexts. Moreover, exploring public-private partnerships could lower the financial barriers for smallholders by introducing innovative financing models, such as shared pyrolysis facilities or co-investment schemes. Finally, studies on other agricultural biomass inputs, such as coffee, corn, and nut shells for biochar production can prove crucial to diversifying the global biochar scene.

Ultimately, with targeted research and supportive policy frameworks, biochar can play a transformative role in Brazil's agricultural landscape, driving both economic and environmental benefits.

Table 6 details the final ranking of studied scenarios based on analysis of the Biochar Economic Viability Model.

**Table 6.** Ranking of studied scenarios based on an analysis of the Biochar Economic Viability Model

| Rank | Scenario | Comment |
| --- | --- | --- |
| 1 | Large B | Viable given investment availability |
| 2 | Large A | Viable given investment availability |
| 3 | Medium B | Viable given investment availability |
| 4 | Small B | Viable, low IRR for high investment |
| 5 | Medium A | Barely viable, easily rendered unviable |
| 6 | Small A | Not viable |

In conclusion, an evaluation of sugarcane bagasse-based biochar' economically viability reveals that sugarcane bagasse-based biochar can in fact, under the right conditions, be a viable option for Brazil's most notable agricultural and environmental initiatives.

**Declaration of Competing Interest**

The author declares that they have no known competing financial interests or personal relationships that could have appeared to influence the work reported in this paper.

**Acknowledgements**

The author is grateful for the opportunity and all the support, insights, and expertise provided by PSR Energy Consulting and Analytics, Rio de Janeiro, Brazil.

S. Nosenzo                                              Evaluating the Economic Viability of Sugarcane Bagasse-Based Biochar353-360. https://doi.org/10.1590/s0103-90162013000500010

Chang, Y., Rossi, L., Zotarelli, L., Gao, B., Shahid, M. A., & Sarkhosh, A. (2021). Biochar improves soil physical characteristics and strengthens root architecture in Muscadine grape (Vitis rotundifolia L.). *Chemical and Biological Technologies in Agriculture*, *8*. https://doi.org/10.1186/s40538-020-00204-5

Cherubin, M. R., Oliveira, D. M. D. S., Feigl, B. J., Pimentel, L. G., Lisboa, I. P. L., Gmach, M. R., Varanda, L. L., Morais, M. C., Satiro, L. S., Popin, G. V., Paiva, S. R. de, Santos, A. K. B. D., Vasconcelos, A. L. S. de, Melo, P. L. A. de, Cerri, C. E. P., & Cerri, C. C. (2018). Crop residue harvest for bioenergy production and its implications on soil functioning and plant growth: A review. *Scientia Agricola*, *75*(3). https://doi.org/10.1590/1678-992X-2016-0459.

Clauser, N. M., Gutiérrez, S., Area, M. C., Felissia, F. E., & Vallejos, M. C. (2018). Alternatives of Small-Scale Biorefineries for the Integrated Production of Xylitol from Sugarcane Bagasse. *Journal of Renewable Materials*, *6*(2). https://doi.org/10.7569/JRM.2017.634145

de A. Sousa, J. G., Cherubin, M. R., Cerri, C. E. P., Cerri, C. C., & Feigl, B. J. (2017). Sugar cane straw left in the field during harvest: Decomposition dynamics and composition changes. *Soil Research*, *55*(8), 758. https://doi.org/10.1071/sr16310

Financial Assumptions. (2021, September). In *EPA*. EPA. Retrieved July 9, 2024, from https://www.epa.gov/system/files/documents/2021-09/chapter-10-financial-assumptions.pdf

*Focus-Market Readout*. (2024, July 5). Banco Central Do Brasil. Retrieved July 9, 2024, from https://www.bcb.gov.br/en/publications/focusmarketreadout

Galinato, S. P., Yoder, J. K., & Granatstein, D. (2011). The economic value of biochar in crop production and carbon sequestration. *Energy Policy*, *39*(10), 6344-6350. https://doi.org/10.1016/j.enpol.2011.07.035

*Gold Standard Fee Schedule* [Fact sheet]. (2023, July 7). Gold Standard for the Global Goals. Retrieved July 9, 2024, from https://globalgoals.goldstandard.org/fees/

Gonçalves, E. V., Seixas, F. L., de Souza Scandiuzzi Santana, L. R., Scaliante, M. H. N. O., & Gimenes, M. L. (2017). Economic trends for temperature of sugarcane bagasse pyrolysis. *The Canadian Journal of Chemical Engineering*, *95*(7), 1269-1279. https://doi.org/10.1002/cjce.22796

Hammond, J., Shackley, S., Sohi, S., & Brownsort, P. (2011). Prospective life cycle carbon abatement for pyrolysis biochar systems in the UK. *Energy Policy*, *39*(5), 2646-2655. https://doi.org/10.1016/j.enpol.2011.02.033

Hermans, J. (2013). Biochar benefits: An introduction to making biochar. *ReNew: Technology for a Sustainable Future*, *124*, 76-79. https://www.jstor.org/stable/renetechsustfutu.124.76

Homagain, K., Shahi, C., Luckai, N., & Sharma, M. (2015). Life cycle environmental impact assessment of biochar-based bioenergy production and utilization in northwestern ontario, canada. *Journal of Forestry Research*, *26*(4), 799-809. https://doi.org/10.1007/s11676-015-0132-y

Homagain, K., Shahi, C., Luckai, N., & Sharma, M. (2016). Life cycle cost and economic assessment of biochar-based bioenergy production and biochar land application in northwestern ontario, canada. *Forest Ecosystems*, *3*(1). https://doi.org/10.1186/s40663-016-0081-8

Jeffery, S., Abalos, D., Spokas, K., & Verheijen, F. G. A. (2014). Biochar for Environmental Management 2 - Biochar effects on crop yield. 299–324.

Joseph, S., Cowie, A. L., Van Zwieten, L., Bolan, N., Budai, A., Buss, W., Cayuela, M. L., Graber, E. R., Ippolito, J. A., Kuzyakov, Y., Luo, Y., Ok, Y. S., Palansooriya, K. N., Shepherd, J., Stephens, S., Weng, Z., & Lehmann, J. (2021). How biochar works, and when it doesn't: A review of mechanisms controlling soil and plant responses to biochar. *GCB Bioenergy*, *13*(11), 1731-1764. https://doi.org/10.1111/gcbb.12885

Kulyk, N. (2012, May). *Cost-Benefit Analysis of the Biochar Application in the U.S. Cereal Crop Cultivation*. University of Massachusetts Amherst. Retrieved July 10, 2024, from https://www.semanticscholar.org/paper/Cost-Benefit-Analysis-of-the-Biochar-Application-in-Kulyk/097a06c804497314fc4447b828fe8e554a97860d

Kung, C.-C., McCarl, B. A., & Cao, X. (2013). Economics of pyrolysis-based energy production and biochar utilization: A case study in taiwan. *Energy Policy*, *60*, 317-323. https://doi.org/10.1016/j.enpol.2013.05.029

Lal, R. (2005). World crop residues production and implications of its use as a biofuel. *Environment International*, *31*(4), 575-584. https://doi.org/10.1016/j.envint.2004.09.005

Lehmann, J., Pereira da Silva Jr, J., Steiner, C., Nehls, T., Zech, W., & Glaser, B. (2003). Nutrient availability and leaching in an archaeological Anthrosol and a Ferralsol of the Central Amazon basin: fertilizer, manure and charcoal amendments. *Plant and Soil*, *249*(2), 343-357. https://doi.org/10.1023/a:1022833116184
16

S. Nosenzo                                                                                          Evaluating the Economic Viability of Sugarcane Bagasse-Based Biochar